\documentclass[12pt]{article}
\usepackage{epsf}

\begin{document}

\newcommand{\gtrsim}{ \mathop{}_{\textstyle \sim}^{\textstyle >} }
\newcommand{\lesssim}{ \mathop{}_{\textstyle \sim}^{\textstyle <} }

\renewcommand{\thefootnote}{\fnsymbol{footnote}}
\setcounter{footnote}{0}

\begin{titlepage}

\def\thefootnote{\fnsymbol{footnote}}

\begin{center}

\hfill TU-613\\
\hfill hep-ph/0103182\\
\hfill March, 2001\\

\vskip .5in

{\Large \bf
No-Scale Scenarios in the Light of New Measurement of Muon 
Anomalous Magnetic Moment
}

\vskip .45in

{\large
Shinji Komine,\footnote{e-mail: komine@tuhep.phys.tohoku.ac.jp}
Takeo Moroi\footnote{e-mail: moroi@tuhep.phys.tohoku.ac.jp} and 
Masahiro Yamaguchi\footnote{e-mail: yama@tuhep.phys.tohoku.ac.jp}
}

\vskip .45in

{\em
  Department of Physics,  Tohoku University, Sendai 980-8578, Japan
}

\end{center}

\vskip .4in

\begin{abstract}

    Supersymmetric contribution to the muon anomalous magnetic moment
    $a_\mu$ is discussed in the no-scale-type supersymmetry breaking
    scenarios.  Taking the correlation between the supersymmetric
    contributions to $a_\mu$ and $Br(b\rightarrow s\gamma)$, it is
    shown that the precise measurements of these quantities serve an
    important constraint on the relative sign of the gaugino masses;
    combining the 2.6-$\sigma$ deviation in $a_\mu$ from the
    standard-model prediction measured by the E821 experiment and
    $Br(b\rightarrow s\gamma)$ measured by CLEO, the sign of the
    product $M_2M_3$ is strongly preferred to be positive, where $M_2$
    and $M_3$ are $SU(2)_L$ and $SU(3)_C$ gaugino mass parameters,
    respectively.  In particular, no-scale-type models with universal
    gaugino masses are in accord with the two constraints and also
    with the Higgs mass bound.  In addition, it is also shown that
    future improvements in the measurements of $a_\mu$ and
    $Br(b\rightarrow s\gamma)$ may provide serious test of the cases
    with $M_2M_3<0$.

\end{abstract}
\end{titlepage}

\renewcommand{\thepage}{\arabic{page}}
\setcounter{page}{1}
\renewcommand{\thefootnote}{\#\arabic{footnote}}
\setcounter{footnote}{0}

Although supersymmetry (SUSY) is a promising candidate for the
solution to the gauge hierarchy problem, issues of SUSY breaking and
its mediation have been a long standing problem in SUSY model
building.  From phenomenological point of view, the mediation
mechanism is of particular interest: it is what we will be able to
probe in future collider experiments.

While superparticles may be too heavy to be directly produced at
existing colliders, they will affect various physical quantities at
loop levels. Among other things, quantities which are rather
insensitive to assumptions on how to suppress SUSY
flavor-changing-neutral-current (FCNC) contributions include a
measured branching ratio $Br(b \rightarrow s \gamma)$ which is in good
agreement with the standard-model prediction, the lightest Higgs mass
$m_h$ which is now bounded to be $m_h\geq 113.5\ {\rm GeV}$ at LEP200
\cite{LEPHiggs}, and muon $g-2$ or muon anomalous magnetic moment
$a_{\mu}\equiv (g-2)/2$ (\cite{previousg-2} and references therein).

Recently, the Brookhaven E821 experiment reported the new measurement
of the muon anomalous magnetic moment $a_{\mu}(\mbox{E821})= 11\ 659\ 
202\ (14)(6) \times 10^{-10}$ which is 2.6-$\sigma$ away from the
standard-model prediction $a_{\mu}(\mbox{SM})$ \cite{E821}.  If we
require that this anomaly be explained by the loop effects of the SUSY
particles, it imposes non-trivial constraints on the MSSM parameters.
(For recent discussions, see Refs.\ \cite{recentg-2}.) In this paper,
we take this apparent deviation very seriously and consider it due to
superparticle contributions
\begin{eqnarray}
a_{\mu}(\mbox{SUSY}) = 
a_{\mu}(\mbox{E821})-a_{\mu}(\mbox{SM}) = 43(16)\times 10^{-10}.
\end{eqnarray}
This can set upper bounds on superparticle masses, particularly on
left-handed smuon and chargino masses, since the dominant contribution
mostly comes from chargino-sneutrino loop.  Importantly, the SUSY
contribution to the muon $g-2$ is approximately proportional to
$\tan\beta$, which is the ratio of the vacuum expectation values of
the two Higgs bosons in the MSSM, and for not very large
$\tan\beta\lesssim 10$, for e.g., the left-handed smuon is required to
be relatively light, $m_{\tilde \mu}\lesssim 400$ GeV, at 2-$\sigma$
level \cite{KMY}.

The upper bounds on the superparticle masses implied by
$a_{\mu}(\mbox{SUSY})$ have interesting interplay with the other two
quantities.  First, in order to satisfy the Higgs mass bound of 113.5
GeV, large radiative corrections from top and scalar top (stop) loops
are required.  This argument suggests heavy stop quarks, in which case
the large positive radiative corrections to the Higgs mass are
expected. This should be contrasted with the light smuon required by
the SUSY interpretation of $a_{\mu}$.  Thus the preferred pattern of
superparticle mass spectrum is that sleptons are light while squarks
are heavy.

Interplay with $Br(b \rightarrow s \gamma)$ is more subtle.  It is
known that the charged-Higgs contribution in the minimal
supersymmetric standard model (MSSM) is always additive to the
standard-model contribution, while the superparticle contribution can
have either positive or negative sign.  Since the measured branching
ratio is consistent with the standard-model prediction, the
charged-Higgs contribution should be cancelled by the other
contributions.  It is nontrivial whether the two requirements that the
positiveness of $a_{\mu}(\mbox{SUSY})$ and the cancellation in $Br(b
\rightarrow s \gamma)$ are simultaneously satisfied, and this will
serve an important test of the mediation mechanisms of the SUSY
breaking \cite{FenMor}.

In the framework of supergravity, tree-level SUSY breaking scalar
masses are given as curvatures of scalar potential in supergravity
Lagrangian.  These masses are thought to be given at some high-energy
renormalization scale close to the Planck scale or the GUT scale. The
low energy values of the scalar masses are evaluated by
renormalization group (RG) methods. Generally speaking, squarks
acquire large positive masses squared due to RG effects from
gluon-gluino loop while sleptons do not.  Thus the squarks become much
heavier than the sleptons especially when the tree-level masses coming
from the supergravity interactions are small and the gaugino mass
contributions from the RG effects dominate.  As we mentioned above,
this pattern of the superparticle mass spectrum is favored from the
$a_{\mu}$ and the Higgs mass bound.

This argument leads us to consider an interesting scenario where the
SUSY breaking scalar masses vanish at some high energy scale close to
the GUT scale or the Planck scale. Such boundary conditions are called
no-scale boundary conditions (see below for a precise definition)
\cite{noscale}. The purpose of the present paper is to study
supersymmetric standard models with no-scale boundary conditions for
SUSY breaking masses in the light of the new result on the muon
anomalous magnetic moment.  We will identify the preferred region for
$a_{\mu}$, taking into account other constraints such as the Higgs
boson mass bound, superparticle mass bounds from direct searches, and
the branching ratio $Br(b\rightarrow s\gamma)$.  We do not impose the
cosmological requirement that the lightest superparticle (LSP) should
be neutral or constitute a dominant part of the dark matter of the
Universe. This is in fact an appealing requirement, but not necessary:
one can consider, for e.g., $R$-parity violation, light gravitino LSP,
or other candidates for the dark matter.

We will consider the following K\"ahler potential in the supergravity:
\begin{equation}
   K = -3 \ln [f(z,z^*) +g(y,y^*)],  
\label{eq:Kahler-no-scale}
\end{equation}
where $f$ is a function of hidden sector fields $z$ responsible for
SUSY breaking and $g$ is a function of observable sector fields $y$
including the MSSM. Here the reduced Planck scale has been set to
unity. It turns out that SUSY breaking scalar masses as well as
trilinear scalar couplings (so-called $A$-terms) vanish at the energy
scale where the boundary conditions are given, as the vacuum
expectation value of the scalar potential ({\it i.e.} the cosmological
constant) vanishes.  The form (\ref{eq:Kahler-no-scale}) can be traced
back to the original no-scale model with a non-compact global symmetry
\cite{noscale}.  The separation of the hidden sector from the
observable sector in the form (\ref{eq:Kahler-no-scale}) was argued to
be natural since it is obtained by the separation of the two sectors
in the superspace density in the supergravity Lagrangian \cite{IKYY}.
Interestingly it is geometrically realized in the setting of two
separated 3-branes in 5-dimensional total spacetime (bulk), namely the
hidden sector lives on one brane and the observable sector lives on
another \cite{sequesterd}. Gaugino masses can arise if, for instance,
the gauge multiplets in the MSSM sector live in the bulk and the
gauginos couple to the hidden sector brane directly (gaugino
mediation) \cite{gauginoMSB}.  The no-scale boundary conditions should
be set at the scale of the inverse of the 5-th dimension's length,
which can be somewhat arbitrary.

The boundary conditions we will consider in this paper are therefore
(i) vanishing scalar masses, (ii) vanishing $A$-terms, and (iii)
non-vanishing gaugino masses $M_i^0$ ($i=1$, 2, 3) for three gauge
groups $U(1)_Y$, $SU(2)_L$ and $SU(3)_C$.  In addition, we assume that
the higgsino mass parameter $\mu$ and Higgs mass mixing parameter
(so-called $B$-parameter) are free parameters in our framework.  The
boundary conditions are given at some high energy scale $M_{\rm bc}$
which we assume to be at the GUT scale or above.  For simplicity we
take all the mass parameters to be real.

The superparticle mass spectrum is obtained by solving the RG
equations.  Specifically we consider particle contents of the MSSM.
We impose the electroweak symmetry breaking condition to reproduce the
correct $Z$ boson mass for given $\tan\beta$.  This constrains, for
instance, $\mu$ and $B$ in our setting.  Thus the masses of the
superparticles and Higgs bosons depend on $M_i^0$ ($i=$1, 2, 3),
$\tan\beta$ and $M_{\rm bc}$.

Let us first consider the case where the gaugino masses are universal
at the energy scale $M_{\rm bc}$. In this case due to RG effects their
values at electroweak scale become approximately $M_1 :M_2 :M_3
\approx 1:2:6$.  As we will see shortly, the relative sign of the
gaugino masses is important when discussing the correlation between
$a_{\mu}$ and $Br(b \rightarrow s \gamma)$.

\begin{figure}[t]
    \centerline{\epsfxsize=0.75\textwidth\epsfbox{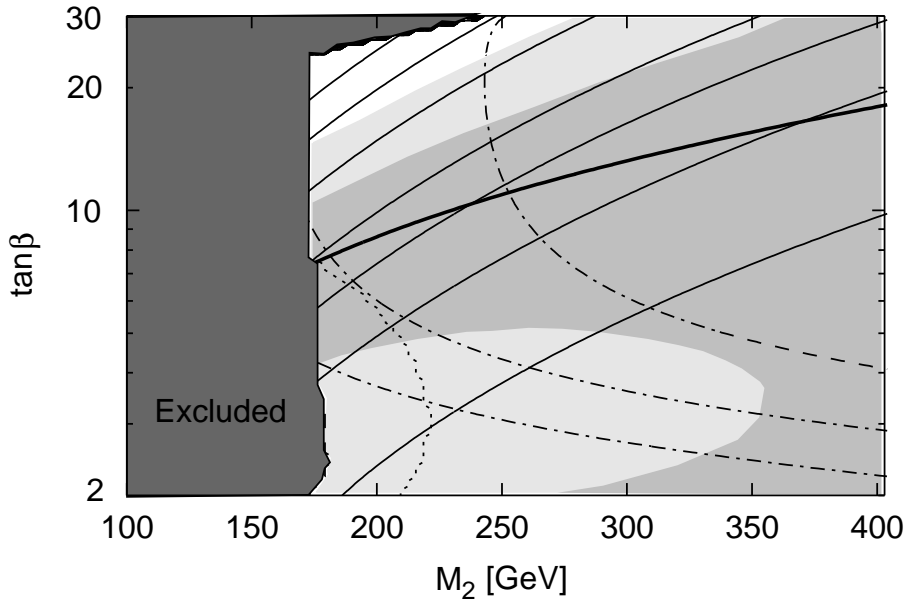}}
    \caption{$a_{\mu}(\mbox{SUSY})$ on the $M_2$ vs.\  $\tan\beta$
    plane for $M_{\rm bc}=2\times 10^{16}\ {\rm GeV}$.  The thin solid
    lines are for $a_{\mu}(\mbox{SUSY})=10$, 20, 30, 40, 60, 80, and
    100 in units of $10^{-10}$ (from below).  The thick solid line is
    the contour for $Br(b\rightarrow s\gamma)=3.15\times 10^{-4}$, the
    center value measured by CLEO, and the darkly- and lightly-shaded
    regions are the parameter space consistent with $Br(b\rightarrow
    s\gamma)=(3.15 \pm 0.54)\times 10^{-4}$ and $2.0\times 10^{-4}<
    Br(b\rightarrow s\gamma) <4.5\times 10^{-4}$, respectively.  The
    contours of the constant Higgs mass ($m_h=110$, 115 and 120 GeV,
    from below) are shown in the dash-dotted lines.  The dotted line
    represents the contour of $m_{\tilde{\tau}_1}=m_{\chi_1}$; on the
    left-handed side of the dotted line,
    $m_{\tilde{\tau}_1}<m_{\chi_1}$ is realized.  Regions with small
    $M_2$ and large $\tan\beta$ are excluded because of negative
    slepton mass and slepton mass smaller than the LEP bound,
    respectively; such regions are also shaded.}
\label{fig:bc=GUT}
\end{figure}

In Fig.\ \ref{fig:bc=GUT}, contours of the constant
$a_{\mu}(\mbox{SUSY})$ are shown in the $M_2$ vs.\ $\tan\beta$ plane
for the case of the universal gaugino masses.  Here, the boundary
conditions are given at $M_{\rm bc} =2\times 10^{16}$ GeV, and we take
$\mu >0$ (and $M_i >0$) so that the SUSY loops give positive
contributions to $a_{\mu}$.  In the same figure, we also present
several other constraints, which were already discussed in
Ref. \cite{Komine}.  First, we plot the contours of the constant $m_h$
in the dash-dotted lines.  In addition, we also show the parameter
region consistent with the measured $Br(b\rightarrow s\gamma)$.  In
this analysis, we use the branching ratio measured by the CLEO
experiment \cite{CLEO}:
\begin{eqnarray}
    Br(b\rightarrow s\gamma) = (3.15 \pm 0.35 \pm 0.32 \pm 0.26)
    \times 10^{-4},
\end{eqnarray}
where the first (second, third) uncertainty is statistical
(systematic, model dependence).  Adding the uncertainties in
quadrature, we obtained the 1-$\sigma$ bound as $Br(b\rightarrow
s\gamma)=(3.15 \pm 0.54)\times 10^{-4}$.  Parameter region consistent
with this 1-$\sigma$ bound is shown by the darkly-shaded region in
Fig.\ \ref{fig:bc=GUT}.  In addition, the lightly-shaded region in
Fig.\ \ref{fig:bc=GUT} is the 95 \% C.L. limit $2.0\times 10^{-4}<
Br(b\rightarrow s\gamma) <4.5\times 10^{-4}$ given by CLEO
\cite{CLEO}.  Remarkably one can find that the parameter region
favored by the muon magnetic moment measurement is completely
consistent with the allowed region by the other constraints.  Notice
that, even after imposing the Higgs mass bound of $m_h\geq 113.5\ {\rm
GeV}$, wide parameter region remains.

In the no-scale-type models, the right-handed slepton masses are
primarily from the $U(1)_Y$ gaugino mass.  As a result, adopting the
GUT relation among the gaugino masses, the right-handed sleptons
become relatively light in this framework.  This fact introduces other
constraint on the no-scale-type models.  Most importantly, the
negative search for the sleptons at the LEP experiment sets a lower
bound on $M_2$ since the slepton masses are increasing functions of
$M_2$.  In Fig.\ \ref{fig:bc=GUT}, this lower bound on $M_2$ is also
shown.  In fact, in some case, this bound is from the right-handed
smuon, since the lightest stau mass $m_{\tilde{\tau}_1}$ is so close
to the lightest neutralino mass $m_{\chi_1}$ that the decay products
from the stau cannot be seen in the detector.  In addition, it is
well-known that, when $M_{\rm bc}\sim 10^{16}$ GeV, the right-handed
stau may become lighter than the lightest neutralino and may become
the lightest superparticle in the MSSM sector \cite{IKYY}.  Such a
case should be cosmologically ruled out if the stau does not decay
\cite{KudoYama}.  In Fig.\ \ref{fig:bc=GUT}, we also indicate the
region where the stau becomes lighter than the lightest neutralino.
Even after including these two constraints, we find that there is
still a region in the parameter space which satisfies all the
constraints.

Now, we would like to comment on the correlation between
$a_{\mu}(\mbox{SUSY})$ and $Br(b \rightarrow s \gamma)$. As was
announced earlier, the relative sign of the gluino mass and the Wino
mass plays an essential role.  The chargino-sneutrino loop diagram
dominates the SUSY contribution to the muon $g-2$, and the sign of
$a_{\mu}(\mbox{SUSY})$ is correlated with that of $M_2\mu$; the muon
magnetic moment receive positive and preferred contribution when $M_2
\mu>0$. On the other hand, as for $Br(b\rightarrow s\gamma)$, the SUSY
contribution is mainly from chargino-stop loop diagram and its sign is
correlated to the sign of $A_t\mu$ (where $A_t$ is the trilinear
scalar coupling of the stop sector).  Since the sign of $A_t$ is
essentially controlled by that of the gluino mass $M_3$ through RG
effects, the SUSY contribution to $Br(b\rightarrow s\gamma)$
interferes with the contributions from the standard model and the
charged-Higgs sector either constructively or destructively depending
on the sign of $M_3$.  The measured $Br(b\rightarrow s\gamma)$ is in
good agreement with the standard-model prediction, and hence the
negative interference is favored. It turns out that this corresponds
to $M_3 \mu >0$.  Thus, in the case of the universal gaugino mass
where the sign of $M_2$ and $M_3$ is the same, one can choose the sign
which is favored by the both measurements simultaneously.

\begin{figure}
    \centerline{\epsfxsize=0.75\textwidth\epsfbox{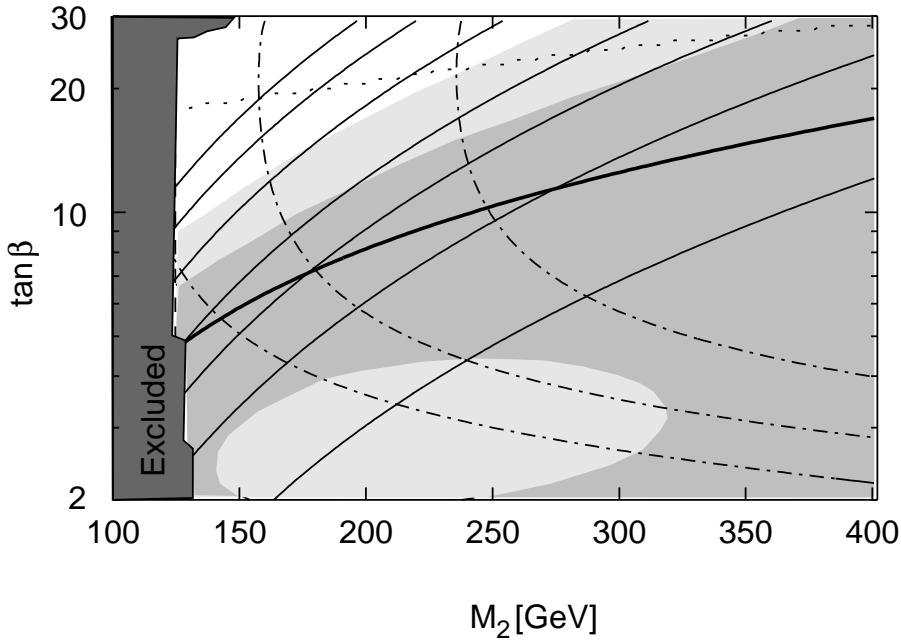}}
    \caption{Same as Fig.\ \ref{fig:bc=GUT} except $M_{\rm bc} = 10^{17}$
    GeV.  The excluded region at around $10\lesssim\tan\beta 
    \lesssim 20$ and
    $130\ {\rm GeV}\lesssim M_2\lesssim 150 {\rm GeV}$ is due to the
    negative search for the right-handed stau.}
\label{fig:bc=10^17}
\end{figure}

Next, let us consider a case with higher $M_{\rm bc}$.  In Fig.\ 
\ref{fig:bc=10^17}, we show the same figure as Fig.\ \ref{fig:bc=GUT}
except that the RG flow starts at $M_{\rm bc} = 10^{17}$ GeV. Above
the GUT scale, we use the RG equations for the minimal SUSY $SU(5)$
model. Superparticle masses are affected by the RG effects above the
GUT scale.  In particular, above the GUT scale, the right-handed
sleptons receive large (positive) mass corrections which are
proportional to the gaugino mass because they belong to {\bf 10}
representation in $SU(5)$ and hence the gaugino loop contribution has
a large group factor.  Other superparticle masses are not modified
drastically. It is well-known that in this case the cosmological
constraint, if we impose, becomes substantially weakend as in a wide
region of the parameter space the stau is heavier than the lightest
neutralino and the neutralino LSP is guaranteed \cite{aboveGUT}.  On
the contrary, the bounds from $b\rightarrow s\gamma$ and from the
Higgs boson mass are almost unchanged when compared with Fig.\ 
\ref{fig:bc=GUT}. This is because these bounds are correlated to the
squark masses, which do not significantly change when one includes the
RG effects above the GUT scale.

As for $a_{\mu}$, the region favored by $a_{\mu}$ is not changed
drastically.  This can be understood if one recalls that the dominant
contribution comes from the chargino-sneutrino loop where the
left-handed slepton mass plays an important role. In $SU(5)$, the
left-handed sleptons are in ${\bf 5^*}$ representation and below the
GUT scale they are $SU(2)_L$ doublets, and thus the RG effect from the
GUT region is not very important.

Thus our main conclusion is that the region favored by $a_{\mu}$ is
completely overlapped with the region preferred by $Br(b\rightarrow
s\gamma)$.  This conclusion remains unchanged even when the RG flow
starts above the GUT scale.

\begin{figure}
    \centerline
    {\epsfxsize=0.75\textwidth\epsfbox{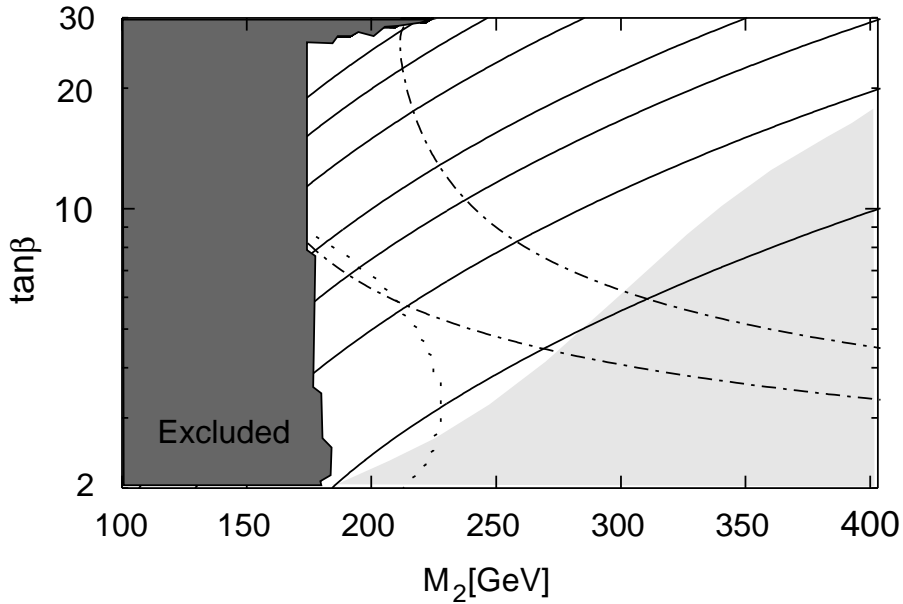}}
    \caption{Same as Fig.\ \ref{fig:bc=GUT} except $M_1^0=M_2^0= -M_3^0$.}
\label{fig:M3<0}
\end{figure}

In order to demonstrate the importance of the relative sign of $M_2$
and $M_3$, we consider the case where gluino mass has an opposite sign
to the other two gaugino masses; namely we take $M_1^0=M_2^0= -M_3^0$
with $M_{\rm bc}=2\times 10^{16}\ {\rm GeV}$, which results in $M_1 :
M_2 :M_3 \approx 1 : 2 : -6$ at the electroweak scale. (For
phenomenological discussions on the case of non-universal gaugino
masses, see Ref.\ \cite{KomYam,Komine}.)  The result is shown in Fig.\ 
\ref{fig:M3<0}.  One can see that the overlap region preferred by both
$a_{\mu}$ and $Br(b\rightarrow s\gamma)$ gets substantially reduced.
This is because, in this case, taking $M_2\mu>0$ for
$a_{\mu}(\mbox{SUSY})>0$, the SUSY contribution to $Br(b\rightarrow
s\gamma)$ constructively interferes with standard-model and
charged-Higgs contributions.  As a result, in order to suppress the
effects of the charged-Higgs and chargino-stop loops, mass scale of
the superparticles is required to be quite high and hence the region
with small $M_2$ is disfavored in Fig.\ \ref{fig:M3<0}.  Thus, unlike
the universal gaugino mass case, the {\em correct} sign of $M_2$ from
$a_{\mu}$ now gives the {\em wrong} sign of $M_{3}$ for $Br(b
\rightarrow s \gamma)$ in this case.  Besides, the Higgs mass bound
constrains more severely than the previous case.  This can be
understood as follows. The left-right mixing mass in the stop sector
has the form $m_t(A_t+\mu\cot\beta)$. In the present case where
$M_3\mu<0$ the two terms tend to cancel each other. Thus the mixing
term becomes small, reducing the radiative corrections from the
stop-top sector to the Higgs boson mass.  Combining the three
constraints, preferred region remains only when the Wino mass is
large, $M_2\gtrsim 250$ GeV. If we further impose the cosmological
constraint that the LSP should be neutral, then the preferred region
completely disappears.

\begin{figure}
    \centerline
    {\epsfxsize=0.75\textwidth\epsfbox{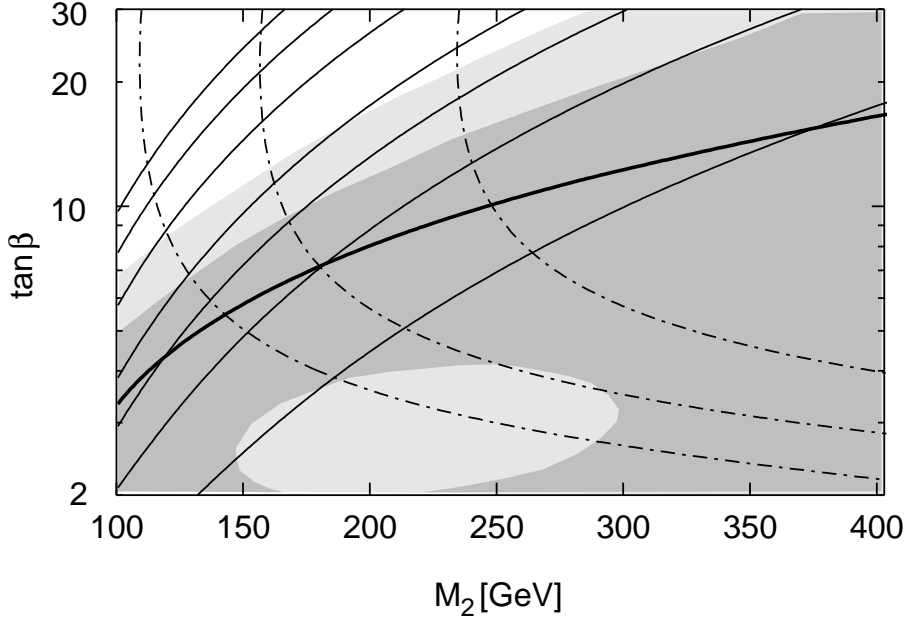}}
    \caption{Same as Fig.\ \ref{fig:bc=GUT} except 
    $M_1^0:M_2^0:M_3^0=3:1:1$.}
    \label{fig:WinoLSP}
\end{figure}

Let us next briefly discuss other examples of non-universal gaugino
masses.  When $M_1$ is much larger than $M_2$, the Wino can be the
LSP. The result with $M_1^0 : M_2^0 :M_3^0=3:1:1$ is shown in
Fig.\ref{fig:WinoLSP}. In this case, the right-handed slepton masses
are significantly pushed up since they originate to the $U(1)_Y$
gaugino mass.  Consequently, the slepton mass bounds disappear (see
Fig.\ref{fig:WinoLSP}).  On the contrary, masses of other sfermions
are insensitive to the change of $M_1$ since they are from $SU(2)_L$
and/or $SU(3)_C$ gaugino mass parameters.  As a result, one finds that
the constraints on $a_{\mu}$, $Br(b \rightarrow s \gamma)$ and the
Higgs mass do not differ from the case of the universal gaugino mass
(see Fig.\ \ref{fig:bc=GUT}).

Another example is the case with smaller $M_3/M_2$.  In this case, for
a fixed value of $M_2$, the RG effects on the squrak masses from the
gluon-gluino loop becomes smaller and hence lighter squarks (in
particular, lighter stops) are realized.  This fact has an important
implication; if the stops are light, the radiative corrections to the
Higgs mass is suppressed.  Thus, in order to evade the Higgs mass
bound, overall mass scale is required to be pushed up.  Then, the
sleptons also become heavier and the SUSY contribution to the muon
$g-2$ is suppressed.  Therefore, for small $M_3/M_2$, constraints on
the parameter space are severer.  To demonstrate this point, we
considered an example with $M_1^0 : M_2^0 : M_3^0=1:1:1/2$.  We
checked that, in this case, the Higgs mass bound severely constrains
the parameter space.  Combining the 2-$\sigma$ constraints on
$a_{\mu}(\mbox{SUSY})$ and $Br(b\rightarrow s\gamma)$ with the Higgs
mass bound $m_h\gtrsim 113.5\ {\rm GeV}$, we found that the parameter
region with $\tan\beta\lesssim 7$ and $M_2\lesssim 280\ {\rm GeV}$ is
excluded.  Furthermore, if the 1-$\sigma$ bound on
$a_{\mu}(\mbox{SUSY})$ {\em or} $Br(b\rightarrow s\gamma)$ is used,
there is no allowed region in the parameter space.  Therefore, among
the models with non-universal gaugino masses, scenarios with lighter
gluino are severely constrained by the combined informations about
$a_{\mu}(\mbox{SUSY})$, $m_h$ and $Br(b\rightarrow s\gamma)$.  Notice
that there is another consequence of the lighter gluino.  If $M_3$ is
small, RG effects to the Higgs mass become smaller and hence the
$\mu$-parameter determined by the radiative electroweak symmetry
breaking condition becomes smaller.  In this case, higgsino
contamination in the lightest neutralino is larger than the universal
gaugino mass case.

In this paper, we considered the case of no-scale boundary conditions.
We should stress here the interesting interplay between $a_{\mu}$ and
$Br(b \rightarrow s \gamma)$ we observed in this paper is very generic
and will be seen in many other classes of SUSY breaking models.

Finally, let us discuss possible improvements of the constraints in
the future.  A new result is expected from the E821 experiment with a
improved statistics.  If the error in $a_{\mu}(\mbox{E821})$ is
reduced by a factor of 2 {\em without changing the center value},
$a_{\mu}(\mbox{SUSY})$ is within the range of $(27-59)\times 10^{-10}$
at the 1-$\sigma$ level.  In this case, the parameter space of the
no-scale model is severely constrained.  Importantly, when the gaugino
masses obey the GUT relation, then even the region with
$a_{\mu}(\mbox{SUSY})=(27-59)\times 10^{-10}$ is consistent with the
constraints from $Br(b\rightarrow s\gamma)$ and the Higgs mass (see
Figs.\ \ref{fig:bc=GUT} and \ref{fig:bc=10^17}).  On the contrary, if
$M_2$ and $M_3$ has opposite sign, then the parameter space is
severely constrained.  In the case with $M_1^0=M_2^0=-M_3^0$, for
e.g., no allowed region remains even if we adopt the 95 \%
C.L. constraint from $Br(b\rightarrow s\gamma)$.  In addition, more
precise measurements of $Br(b\rightarrow s\gamma)$ has an important
impact to constrain the scenario of the SUSY breaking, in particular
if combined with the muon $g-2$ anomaly.  Therefore, improvements of
the statistics in the measurement of the muon $g-2$ \rm{as well as
$Br(b \rightarrow s \gamma)$} will provide interesting and important
informations in studying the scenarios of the SUSY breaking.

{\em Note Added:} In our analysis, we calculated the lightest Higgs
mass using the one-loop effective potential given in Ref.\ 
\cite{DreNoj}.  In Ref.\ \cite{CarHabHei}, the lightest Higgs mass is
calculated taking account of two-loop contributions, and it is pointed
out that the lightest Higgs mass tends to decrease due to the two-loop
effect.  The authors would like thank M.M.\ Nojiri for useful
conversations.

{\em Acknowledgment:} This work was supported in part by the
Grant-in-aid from the Ministry of Education, Culture, Sports, Science
and Technology, Japan, priority area (\#707) ``Supersymmetry and
unified theory of elementary particles,'' and in part by the
Grants-in-aid No.11640246 and No.12047201.


\begin{thebibliography}{100}

\bibitem{LEPHiggs} P. Igo-Kemenes, talk given at the LEPC open session
    (November, 2000),
    http://lephiggs.web.cern.ch/LEPHIGGS/talks/index.html.

\bibitem{previousg-2}
    J.L. Lopez, D.V. Nanopoulos and X. Wang, 
    Phys.\ Rev.\ D49 (1994) 366;
    U. Chattopadhyay and P. Nath,
    Phys.\ Rev.\ D53 (1996) 1648;
    T. Moroi, 
    Phys.\ Rev.\ D53 (1996) 6565; Erratum-ibid, D56 (1997) 4424;
    M. Carena, G.F. Giudice and C.E.M. Wagner, 
    Phys.\ Lett.\ B390 (1997) 234;
    J.L. Feng and T. Moroi,  
    Phys.\ Rev.\ D61 (2000) 095004;
    T. Goto, Y. Okada and Y. Shimizu, hep-ph/9908499;
    T. Blazek, hep-ph/992460;
    U. Chattopadhyay, D.K. Ghosh and S. Roy, 
    Phys.\ Rev.\ D62 (2000) 115001;
    M. Drees, Y.G. Kim, T. Kobayashi and M.M. Nojiri, 
    hep-ph/0011359.

\bibitem{E821}
    H.N. Brown et al.\ (Muon ($g-2$) Collaboration), 
    hep-ex/0102017.

\bibitem{recentg-2}
    A. Czarnecki and W.J. Marciano, hep-ph/0102122;
    L. Everett, G.L. Kane, S. Rigolin and L.-T. Wang, hep-ph/0102145;
    J.L. Feng and K.T. Matchev, hep-ph/0102146;
    E.A. Baltz and P. Gondolo, hep-ph/0102147;
    U. Chattopadhyay and P. Nath, hep-ph/0102157;
    S. Komine, T. Moroi and M. Yamaguchi, hep-ph/0102204;
    J. Ellis, D.V. Nanopoulos and K.A. Olive, hep-ph/0102331;
    J. Hisano and K. Tobe, hep-ph/0102315;
    R. Arnowitt, B. Dutta, B. Hu, Y. Santoso,
    hep-ph/0102344;
    K. Choi, K. Hwang, S.K. Kang, K.Y. Lee and W.Y. Song,
    hep-ph/0103048;
    J.E. Kim, B. Kyae and H.M. Lee, hep-ph/0103054;
    S. Martin and J.D. Wells, hep-ph/0103067.
    
\bibitem{KMY}
    S. Komine, T. Moroi and M. Yamaguchi, in Ref.\ \cite{recentg-2}.
    
\bibitem{FenMor}
    J.L. Feng and T. Moroi,  in Ref.\ \cite{previousg-2};
    K.T. Mahanthappa and S. Oh,
    Phys.\ Rev.\ D62 (2000) 015012.
    
\bibitem{noscale}
    J. Ellis, C. Kounnas and D.V. Nanopoulos,
    Nucl.\ Phys.\ B247 (1984) 373;
    J. Ellis, A.B. Lahanas, D.V. Nanopoulos and K. Tamvakis,
    Phys.\ Lett.\ B134 (1984) 429.
    
\bibitem{IKYY}
    K. Inoue, M. Kawasaki, M. Yamaguchi and T. Yanagida,
    Phys.\ Rev.\ D45 (1992) 328.
    
\bibitem{sequesterd}
    L. Randall and R. Sundrum, Nucl.\ Phys.\ B557 (1999) 79;
    M.A. Luty and R. Sundrum, Phys.\ Rev.\ D62 (2000) 035008.
    
\bibitem{gauginoMSB}
    D.E. Kaplan, G.D. Kribs and M. Schmaltz,
    Phys.\ Rev.\ D62 (2000) 035010;
    Z. Chacko, M.A. Luty, A.E. Nelson and E. Ponton,
    JHEP 0001 (2000) 003.
    
\bibitem{Komine}
    S. Komine, hep-ph/0102030.
    
\bibitem{CLEO}
    CLEO collaboration, hep-ex/9908022.
    
\bibitem{KudoYama}
    A. Kudo and M. Yamaguchi, in preparation.
    
\bibitem{aboveGUT}
    Y. Kawamura, H. Murayama and M. Yamaguchi,
    Phys.\ Rev.\ D51 (1995) 1337;
    N. Polonsky and A. Pomarol,
    Phys.\ Rev.\ D51 (1995) 6532.
    
\bibitem{KomYam}
    S. Komine and M. Yamaguchi,
    Phys.\ Rev.\ D63 (2001) 035005.
    
\bibitem{DreNoj}
    M. Drees and M.M. Nojiri,
    Phys.\ Rev.\ D45 (1992) 2482.

\bibitem{CarHabHei}
    M. Carena, H.E. Haber, S. Heinemeyer, W. Hollik, C.E.M. Wagner and 
    G. Weiglein,
    Nucl.\ Phys.\ B580 (2000) 29.

\end{thebibliography}
\end{document}